\documentclass[aps,prl,twocolumn,showpacs,preprintnumbers,amsmath,amssymb,superscriptaddress]{revtex4}
\usepackage{mathptm}  
\usepackage{dcolumn}                    
\usepackage{bm}                        
\usepackage{graphicx}
\usepackage{times}
\usepackage{epstopdf}
\begin{document}

%% Math
\newcommand{\Imag}{{\Im\mathrm{m}}}   % Imaginary part 
\newcommand{\Real}{{\mathrm{Re}}}   % Real part
\newcommand{\im}{\mathrm{i}}        % Imaginary unit non-italic
\newcommand{\talpha}{\tilde{\alpha}}
\newcommand{\ve}[1]{{\mathbf{#1}}}

\newcommand{\x}{\lambda}  % holder for plus/minus 1 (\pm 1)
\newcommand{\y}{\rho}     % holder for plus/minus 1 (\pm 1)
\newcommand{\T}{\mathrm{T}}   % Time ordering operator
\newcommand{\Pv}{\mathcal{P}} % Principal value operator
\newcommand{\vk}{\ve{k}} % Vector k
\newcommand{\vp}{\ve{p}} % Vector p

\newcommand{\ml}{\boldsymbol{m}_L} 
\newcommand{\mr}{\boldsymbol{m}_R} 

\newcommand{\N}{\underline{\mathcal{N}}} % Vector k
\newcommand{\Nt}{\underline{\tilde{\mathcal{N}}}} % Vector p
\newcommand{\g}{\underline{\gamma}} % Vector k
\newcommand{\gt}{\underline{\tilde{\gamma}}} % Vector p

\newcommand{\vecr}{\ve{r}} % Vector p
\newcommand{\vq}{\ve{q}} % Vector q
\newcommand{\ca}[2][]{c_{#2}^{\vphantom{\dagger}#1}} % op. c (annihilate) 
\newcommand{\cc}[2][]{c_{#2}^{{\dagger}#1}}          % op. c dagger (create) 
\newcommand{\da}[2][]{d_{#2}^{\vphantom{\dagger}#1}} % op. d (annihilate) 
\newcommand{\dc}[2][]{d_{#2}^{{\dagger}#1}}          % op. d dagger (create) 
\newcommand{\ga}[2][]{\gamma_{#2}^{\vphantom{\dagger}#1}} % op. gamma
\newcommand{\gc}[2][]{\gamma_{#2}^{{\dagger}#1}}          % op. gamma dagger 
\newcommand{\ea}[2][]{\eta_{#2}^{\vphantom{\dagger}#1}} % op. eta
\newcommand{\ec}[2][]{\eta_{#2}^{{\dagger}#1}}          % op. eta dagger 
\newcommand{\su}{\uparrow}    % Make the code more readable...
\newcommand{\sd}{\downarrow}  % Make the code more readable...
\newcommand{\Tkp}[1]{T_{\vk\vp#1}}  % Tunneling matrix element
\newcommand{\muone}{\mu^{(1)}}      % Chem.pot. side one
\newcommand{\mutwo}{\mu^{(2)}}      % Chem.pot. side two
\newcommand{\epsk}{\varepsilon_\vk}
\newcommand{\epsp}{\varepsilon_\vp}
\newcommand{\e}[1]{\mathrm{e}^{#1}}
\newcommand{\dif}{\mathrm{d}} %Rett d i differensial
\newcommand{\diff}[2]{\frac{\dif #1}{\dif #2}}%Derivert
\newcommand{\pdiff}[2]{\frac{\partial #1}{\partial #2}}%Derivert
\newcommand{\mean}[1]{\langle#1\rangle}
\newcommand{\abs}[1]{|#1|}
\newcommand{\abss}[1]{|#1|^2}
\newcommand{\Sk}[1][\vk]{\ve{S}_{#1}}
\newcommand{\pauli}[1][\alpha\beta]{\boldsymbol{\sigma}_{#1}^{\vphantom{\dagger}}}

%% Text
\newcommand{\eq}{Eq.}%No extra space when used with reftex (->auto ~)
\newcommand{\eqs}{Eqs.}%No extra space when used with reftex (->auto ~)
\newcommand{\cf}{\textit{cf. }}%adv : that is to say; in other words
\newcommand{\ie}{\textit{i.e. }}%adv : that is to say; in other words
\newcommand{\eg}{\textit{e.g. }}%[syn: f.eks., for example, for instance]
\newcommand{\etal}{\emph{et al.}}
\def\i{\mathrm{i}}

\title{Supercurrent-Induced Magnetization Dynamics}

\author{Jacob Linder}
\affiliation{Department of Physics, Norwegian University of
Science and Technology, N-7491 Trondheim, Norway}

\author{Takehito Yokoyama}
\affiliation{Department of Physics, Tokyo Institute of Technology, 2-12-1 Ookayama, Meguro-ku, Tokyo 152-8551, Japan}

\date{\today}

\begin{abstract}

We investigate supercurrent-induced magnetization dynamics in a Josephson junction with two misaligned ferromagnetic layers, and demonstrate a variety of effects by solving numerically the Landau-Lifshitz-Gilbert equation. In particular, we demonstrate the possibility to obtain supercurrent-induced magnetization switching for an experimentally feasible set of parameters, 
and clarify the favorable condition for the realization of magnetization reversal. 
These results constitute a superconducting analogue to conventional current-induced magnetization dynamics and indicate how spin-triplet supercurrents may be utilized for practical purposes in spintronics. 

\end{abstract}
\maketitle

\textit{Introduction}. The interplay between superconducting and ferromagnetic order is presently generating much interest in a variety of research communities \cite{bergeret_rmp_05}. Besides the obvious interest from a fundamental physics point of view, a major part of the allure of superconductor$\mid$ferromagnet (S$\mid$F) hybrid structures is the prospect of combining the spin-polarized charge carriers present in ferromagnets with the dissipationless flow of a current offered by the superconducting environment. By tailoring the desired properties of a hybrid S$\mid$F system on a nanometer scale, this interplay opens up new perspectives within spin-polarized transport.

Closely related to the transport of spin is the phenomenon of current-induced magnetization dynamics \cite{tserkovnyak_rmp_05}. The general principle is that a spin-polarized current injected into a ferromagnetic layer can act upon the magnetization of that layer via a torque and thus induce magnetization dynamics \cite{slonczewski_jmmm_96}. 
Previous works in this field have considered mainly non-equilibrium spin accumulation via quasiparticle spin-injection from a ferromagnet into a superconductor \cite{clarke_prl_72, tedrow_prb_73, yeh_prb_99, takahashi_prl_99, morten_prb_05}. 
The concept of supercurrent-induced magnetization dynamics suggests an interesting venue for combining the seemingly disparate fields of superconductivity and spintronics. 
However,  magnetization dynamics in a Josephson junction has so far been discussed only in a handful of works \cite{waintal_prb_02, zhao_prb_08, maekawa,Houzet,konschelle_prl_09}. In particular, it was demonstrated in Ref. \cite{waintal_prb_02} how a supercurrent would induce an equilibrium exchange interaction between two non-collinear ferromagnets in an S$\mid$F$\mid$N$\mid$F$\mid$S junction (N stands for normal metal). Taking into account the fact that a Josephson current flowing through such an inhomogeneous magnetization profile will have a spin-triplet contribution \cite{bergeret_rmp_05}, such an interaction implies that it should be possible to generate \textit{supercurrent-induced magnetization dynamics} in this type of junction. This would constitute a superconducting analogue to magnetization dynamics in a conventional spin-valve setup. To this date, this remains unexplored in the literature. 

\begin{figure}[t!]
\centering
\resizebox{0.48\textwidth}{!}{
\includegraphics{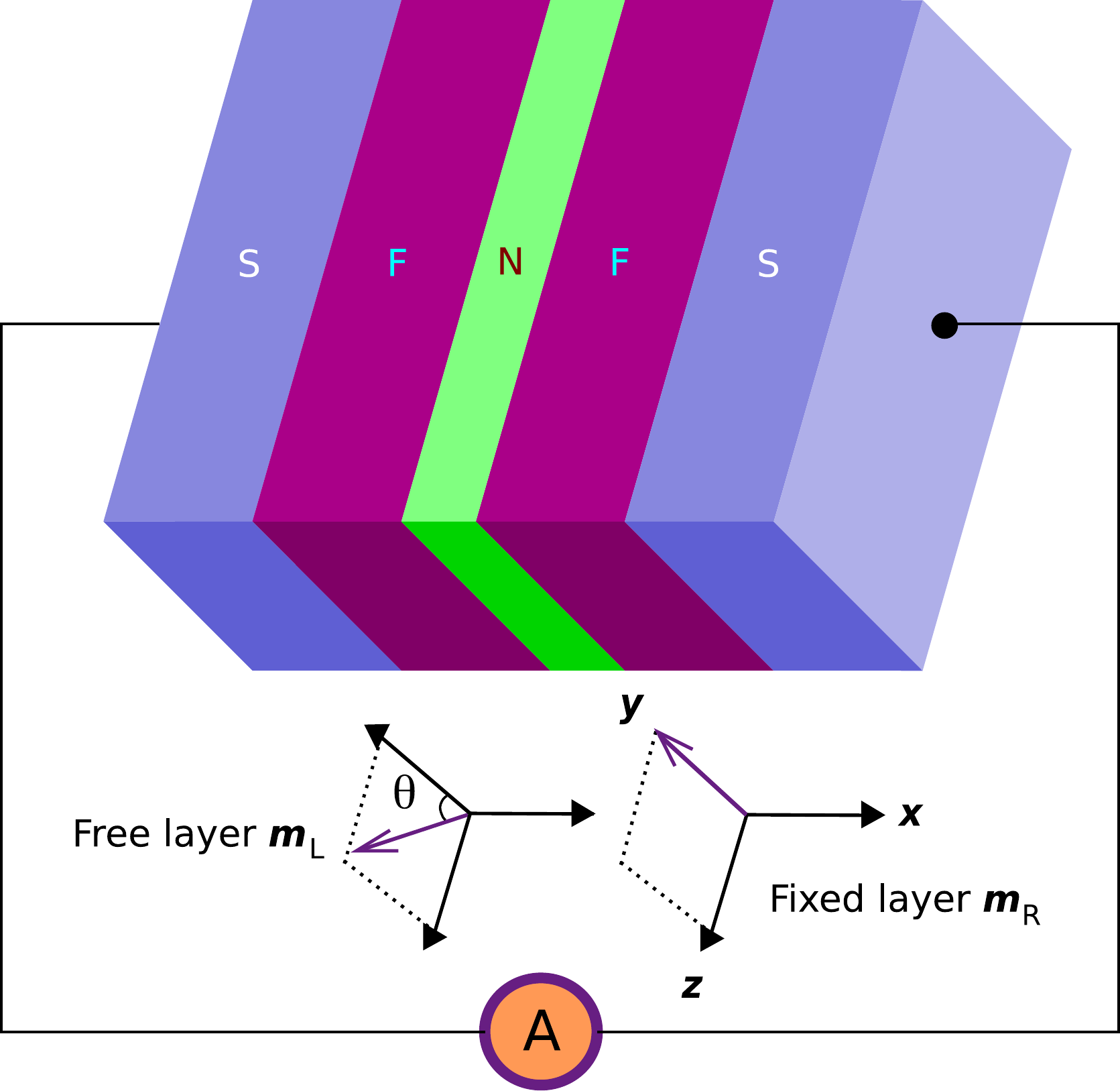}}
\caption{(Color online) The proposed experimental structure. Two ferromagnetic layers separated by a normal spacer are sandwiched between two $s$-wave superconductors. The magnetization directions may be non-aligned and tuned by means of an external field, as long as the exchange coupling between the ferromagnets is sufficiently reduced by the thickness of the normal spacer. By current-biasing this system one may generate a supercurrent which induces magnetization dynamics.}
\label{fig:model} 
\end{figure}

\begin{figure*}[t!]
\centering
\resizebox{0.99\textwidth}{!}{
\includegraphics{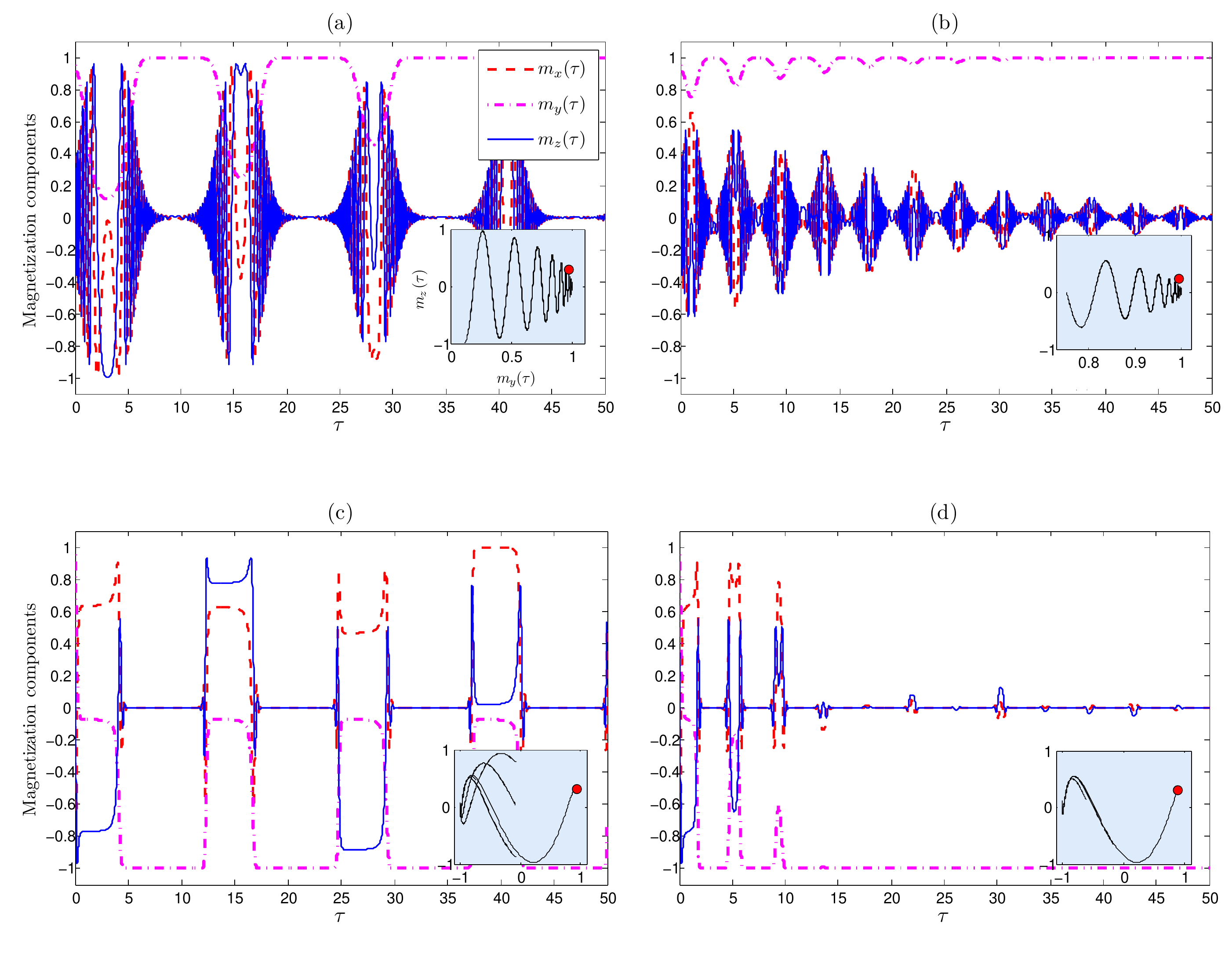}}
\caption{(Color online) The time-evolution of the normalized magnetization components $m_j$. Here, we have set $\zeta=290$ with an initial angle misalignment $\theta_0/\pi=0.1$. We have considered the weak damping regime $\alpha\ll1$ in the upper row, setting (a) $\omega=0.5$ and $\alpha=0.05$ while in (b) $\omega=1.5$ and $\alpha=0.05$. In the lower row, we have considered stronger Gilbert damping and set (c) $\omega=0.5$ and $\alpha=0.5$ while in (d) $\omega=1.5$ and $\alpha=0.5$. The insets display a parametric plot of the $\boldsymbol{\hat{y}}$- and $\boldsymbol{\hat{z}}$-components of the magnzetization with the red circles indicating the point $\tau=0$ (the initial magnetization configuration). The time-span in insets is $\tau\in[0,50]$.}
\label{fig:main} 
\end{figure*}

Motivated by this, we study in this Letter for the first time the magnetization dynamics of a multilayer ferromagnetic Josephson junction when a current bias is applied, based on the Landau-Lifshitz-Gilbert (LLG) equation \cite{llg}. Our main idea is to utilize the spin-triplet nature of the Josephson current, which very recently has been observed experimentally \cite{khaire_prl_10, sprungmann_arxiv_10, robinson_science_10}, in order to induce a torque on the magnetic order parameter and thus generate magnetization dynamics.  The experimentally relevant setup is shown in Fig. \ref{fig:model}: two ferromagnets with different coersive fields are separated by a normal spacer and sandwiched between two conventional $s$-wave superconductors. The coersive fields are such that the magnetic order parameter is hard in one layer, while soft in the other. By application of an external field $\boldsymbol{H}_\text{ext}$, it is thus possible to tune the relative orientation of the local magnetization fields in the two layers. When this junction is current-biased, a supercurrent flows without resistance up to a critical strength. The supercurrent strongly modifies the equilibrium exchange interaction between the ferromagnetic layers and should thus be expected to result in supercurrent-induced magnetization dynamics. We investigate this by solving numerically the LLG equation, which provides the time-evolution of the magnetization components in the soft magnetic layer. As we shall see, a number of interesting opportunities arise in terms of the torque exerted on the soft layer by the Josephson current. We proceed by first establishing the theoretical framework used in the forthcoming analysis and then present our main results.

\textit{Theory}. The magnetization dynamics of the Josephson junction is governed by the LLG equation, which in the free magnetic layer reads:
\begin{align}
\frac{\partial\ml}{\partial t} = -\gamma \ml \times \boldsymbol{H}_\text{eff} + \alpha\Big(\ml \times \frac{\partial \ml}{\partial t}\Big),
\end{align}
where $\gamma$ is the gyromagnetic ratio and $\alpha$ is the damping constant. The effective magnetic field is obtained from the free energy functional via the relation $\boldsymbol{H}_\text{eff} = -(\delta F/\delta \ml)/(\mathcal{V}M_0)$ where $\mathcal{V}$ is the unit volume and $M_0$ is the magnitude of the magnetization. Note that only transverse magnetization dynamics are described by the LLG equation due to the cross-terms coupling to $\ml$, meaning that $|\ml|=1$ is constant in time. To study the magnetization dynamics induced by the supercurrent, we must identify the effective field. To this end, we employ the following phenomenological form \cite{waintal_prb_02} stemming from the exchange interaction generated by the Josephson current:
\begin{align}\label{eq:F}
F_S = \frac{A\Delta_0}{\lambda_F^2}\cos\phi[\mathcal{J}_1(\ml\cdot\mr) + 2\mathcal{J}_2(\ml\cdot\mr)^2 - \mathcal{J}_2],
\end{align}
where $A$ is the unit area, $\lambda_F$ is the Fermi wavelength, and $\Delta_0$ is the gap magnitude. The parameters  $\mathcal{J}_i$ are analogues to the quadratic and biquadratic coupling constants for a magnetic exchange interaction.
In addition to Eq. (\ref{eq:F}), one should also include the anisotropy contribution $F_M$ to the free energy which provides an effective field $\boldsymbol{H}_\text{eff}^M = (K m_y/M_0)\hat{\boldsymbol{y}}$ where $K$ is the anisotropy constant. We have here assumed that the anisotropy axis $\parallel \hat{\boldsymbol{y}}$. After some algebra, one arrives at the final form of the LLG equation in our system:
\begin{align}\label{eq:LLG}
\frac{\partial\ml}{\partial t} = \ml\times\Big( &-\frac{\gamma K m_y}{M_0}\hat{\boldsymbol{y}} + \mr \frac{\gamma\Delta_0\cos\phi(t)}{dM_0\lambda_F^2}[\mathcal{J}_1 + 4\mathcal{J}_2(\ml\cdot\mr)] \notag\\
&+ \alpha \frac{\partial\ml}{\partial t} \Big) 
\end{align}
where $\phi(t) = \omega_Jt$ and $d$ is the thickness of the ferromagnetic layer.
Importantly, we note that the anisotropy contribution proportional to $K$ was \textit{not included} in the effective field used in previous works \cite{waintal_prb_02}. This contribution is nevertheless essential since the supercurrent-induced torque must overcome the anisotropy contribution in order to switch the magnetization orientation. In what follows, we will present a full numerical solution of this equation to investigate the supercurrent-induced magnetization dynamics. To this end, we first establish experimentally relevant values of the parameters in Eq. (\ref{eq:LLG}). The above equation may be cast into a dimensionless form by introducing $\omega_F = \gamma K/M_0$, $\tau = \omega_Ft$, $\omega = \omega_J/\omega_F$, and $\zeta = \Delta_0/(K d\lambda_F^2)$. Here, $\omega_F$ is the ferromagnetic resonance frequency and $\omega_J$ is the Josephson frequency. Employing a realistic estimate \cite{stiles} for transmission probabilities in the F$\mid$N$\mid$F part of the system, one finds \cite{waintal_prb_02} that Eq. (\ref{eq:F}) accounts well for the Josephson current when $\mathcal{J}_1 = 0.007$ and $\mathcal{J}_2 = 0.025$. For a permalloy with weak anisotropy, one may estimate \cite{rusanov_prl_04} $K \simeq 4 \times 10^{-5}$ K {\AA}$^{-3}$. Moreover, we set $\Delta_0=1$ meV, $\lambda_F = 1$ \AA, and $d=10$ nm as standard values \cite{ryazanov_prl_01} for the hybrid structure under consideration. The Josephson frequency $\omega_J$ is typically of order GHz, but may be tuned experimentally. We will therefore consider several choices of $\omega=\omega_J/\omega_F$ and the damping constant $\alpha$ to model a variety of experimentally accessible scenarios. As long as the requirement $\hbar \omega_J \ll T_c$ is fulfilled, i.e. the time-dependent part is a small perturbation, one may consider $\phi(t)$ as a time-dependent external potential in the static expression for the free energy \cite{konschelle_prl_09}. Since typically $\omega_J \sim 1$ $\mu$V, this condition is easily met. The supercurrent-induced magnetization dynamics come into play when the local magnetizations $\{\ml,\mr\}$ are misaligned with an angle $\theta_0 = \theta(t=0)$ to begin with, and vanishes when $\theta_0=\{0,\pi\}$. The magnetization in the right (hard) layer is fixed at $\mr = \boldsymbol{\hat{y}}$.

\textit{Results and Discussion.} Using the parameters discussed above, we now proceed to investigate the resulting magnetization dynamics for both weak damping $(\alpha\ll1)$ and more considerable damping $(\alpha \sim 1)$. Choosing a small initial angle of misalignment $\theta_0/\pi=0.1$, the numerical solution of the LLG-equation is shown in Fig. \ref{fig:main}. The upper row shows the weak-damping regime for two different choices of the frequency ratio ($\omega <1$ and $\omega>1$). The qualitative behavior is similar in (a) and (b): the supercurrent-induced torque induces oscillations of the magnetization components, but is unable to switch the magnetization direction in the soft layer from its original configuration $\ml \simeq \boldsymbol{\hat{y}}$. The oscillations eventually die out and the magnetization orientation of the two layers saturates in a parallel configuration. This happens on a faster time-scale in (b) due to the larger value of $\omega$ used compared to (a). Consider now the case of stronger Gilbert damping shown in the lower row of Fig. \ref{fig:main}. The a.c. Josephson current-induced torque now induces more "violent" oscillations and is able to rotate the magnetization orientation by $\pi$. When the frequency is large enough, exemplified in (d) by $\omega=1.5$, \textit{full magnetization reversal} is achieved and maintained. On a larger time-scale, the same behavior is seen for lower frequencies. In general, we observe in our numerical simulations that the main difference between the weak and strong Gilbert damping regime is that the magnetization dynamics is periodic and oscillating in the former case, whereas it tends to saturate to a fixed orientation in the latter case. This is reasonable as the damping effectively leads to a more rapid decay of the oscillating magnetization dynamics induced by the a.c. current. The results in Fig. \ref{fig:main} suggest that it is possible to generate supercurrent-induced magnetization reversal in the setup shown in Fig. \ref{fig:model}. This phenomenon should be intimately linked with the spin-triplet correlations present in the non-collinear setup considered here since spin-singlet correlations cannot carry torque. Recently, controllable long-range triplet supercurrents have been experimentally observed in Josephson junctions with strong inhomogeneous ferromagnetic layers \cite{khaire_prl_10, sprungmann_arxiv_10, robinson_science_10}. In the present setup, torque carried by the supercurrent does not become long-ranged \cite{sperstad_prb_08} (an additional source of triplet correlations would be required for that purpose \cite{houzet_prb_07}) whereas it is still due to spin-triplet, thus mediating magnetic correlations between the two ferromagnetic layers.

\begin{figure}[t!]
\centering
\resizebox{0.5\textwidth}{!}{
\includegraphics{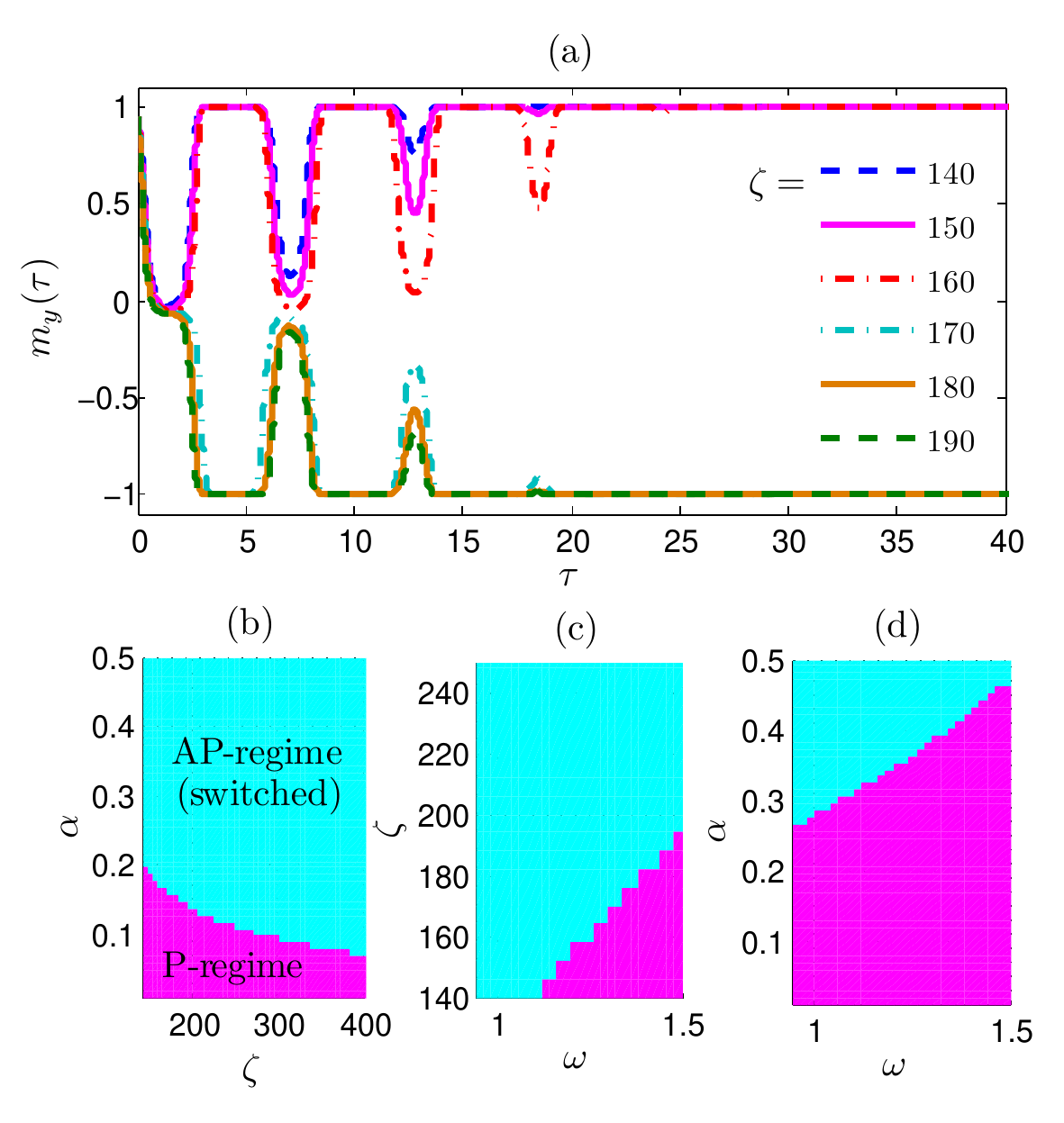}}
\caption{(Color online) (a) The time-evolution of the normalized magnetization component $m_y$. Here, we have set $\omega=1.1$ and $\alpha=0.4$ with an initial angle misalignment $\theta_0/\pi=0.1$. Above a critical value of $\zeta$, permanent switching occurs from $\ml \parallel \boldsymbol{\hat{y}}$ to $\ml \parallel (-\boldsymbol{\hat{y}})$. The phase-diagram for magnetization reversal [from the parallel (P) configuration $\ml \parallel \mr$ to the anti-parallel (AP) configuration $\ml \parallel (-\mr)$] is shown in (b-d). A contour-plot is given for $m_y(t\to\infty)$ with (b) $\omega=0.5$, (c) $\alpha=0.5$, and (d) $\zeta=200$.}
\label{fig:switch} 
\end{figure}

We proceed by investigating the role of the anisotropy preference in the soft magnetic layer. This is modelled by the term proportional to $K$ in Eq. (\ref{eq:LLG}). The parameter $\zeta$ effectively models the relative weight of the anisotropy energy and the Josephson energy related to the presence of superconductivity. In Fig. \ref{fig:switch}(a), we focus on the behavior of the magnetization component $m_y$.
We plot its time-evolution for increasing values of $\zeta$. 
It is seen that supercurrent-induced magnetization reversal occurs for $\zeta > \zeta_c$ with a critical value $\zeta_c$ (for the present parameters, we find $\zeta_c\simeq 163$). The magnetization then saturates, and the magnitude of the oscillations decreases as $\zeta$ becomes larger. In Fig. \ref{fig:switch}(b)-(d), we give a phase-diagram for the magnetization switching by plotting $m_y(t\to\infty)$ in the $\alpha-\zeta$, $\zeta-\omega$, and $\alpha-\omega$ plane. 
In general, the results indicate that the torque generated by the a.c. Josephson effect can reverse the magnetization orientation when the Gilbert damping is non-neglible, the anisotropy contribution is sufficiently weak and Josephson frequency is sufficiently small.  
The phase-diagram in (b) and (d) shows that the magnetization can be trapped in the switched direction when the damping $\alpha$ is sufficiently large, in agreement with the results in Fig. 2.
It should also be noted that the switching becomes less viable as $\omega$ increases as seen from  (c) and (d). When $\omega_J \gg \omega_F$, the torque from the Josephson current oscillates 
very fast compared to the characteristic motion of the magnetization. In this way, the magnetization would experience an averaged torque over many oscillations, which results in small effect due to partial cancellation of the net torque.

\textit{Conclusion.} In summary, we have investigated supercurrent-induced magnetization dynamics in a Josephson junction with two misaligned ferromagnetic layers by solving numerically the Landau-Lifshitz-Gilbert equation. We have demonstrated the possibility to obtain supercurrent-induced magnetization switching for an experimentally realistic parameter set. 
It is clarified that for the realization of magnetization reversal, large Gilbert damping, small anisotropy contribution or a small Josephson frequency are in general favorable.
These results constitute a superconducting analogue to conventional current-induced magnetization dynamics and indicate how spin-triplet supercurrents may be utilized for practical purposes in spintronics.


\begin{thebibliography}{99}

\bibitem{bergeret_rmp_05} F. Bergeret \etal, Rev. Mod. Phys. \textbf{77}, 1321 (2005); A. Buzdin, Rev. Mod. Phys. \textbf{77}, 935 (2005).

\bibitem{tserkovnyak_rmp_05} Y. Tserkovnyak \etal, Rev. Mod. Phys. \textbf{77}, 1375 (2005).

\bibitem{slonczewski_jmmm_96} J. Slonczewski, J. Magn. Magn. Mater. \textbf{159}, L1 (1996).

\bibitem{clarke_prl_72} J. Clarke, Phys. Rev. Lett. \textbf{28}, 1363 (1972).

\bibitem{tedrow_prb_73} P. M. Tedrow and R. Meservey, Phys. Rev. B \textbf{7}, 318 (1973).

\bibitem{yeh_prb_99} N. C. Yeh \etal, Phys. Rev. B \textbf{60}, 10522 (1999).

\bibitem{takahashi_prl_99} S. Takahashi \etal, Phys. Rev. Lett. \textbf{82}, 3911 (1999).
\bibitem{morten_prb_05} J. P. Morten \etal, Phys. Rev. B \textbf{72}, 014510 (2005).

\bibitem{waintal_prb_02} X. Waintal and P. Brouwer, Phys. Rev. B \textbf{65}, 054407 (2002).

\bibitem{maekawa}
S. Takahashi, S. Hikino, M. Mori, J. Martinek, and S. Maekawa, Phys. Rev. Lett. \textbf{99}, 057003 (2007).% S. Hikino, M. Mori, S. Takahashi, S. Maekawa, J. Phys. Soc. Jpn \textbf{77}, 053707 (2008).

\bibitem{zhao_prb_08} E. Zhao and J. Sauls, Phys. Rev. B \textbf{78}, 174511 (2008).

\bibitem{Houzet} M. Houzet, Phys. Rev. Lett. {\bf 101}, 057009 (2008).

\bibitem{konschelle_prl_09} F. Konschelle and A. Buzdin, Phys. Rev. Lett. \textbf{102}, 017001 (2009).

\bibitem{llg} L. D. Landau and E. M. Lifshitz, Phys. Z. Sowjet. \textbf{8}, 153 (1935); T. L. Gilbert, IEEE Trans. Magn. \textbf{40}, 3443 (2004).

\bibitem{rusanov_prl_04} A. Y. Rusanov \etal, Phys. Rev. Lett. \textbf{93}, 057002 (2004).

\bibitem{stiles} M. D. Stiles, J. Appl. Phys. \textbf{79}, 5805 (1996); Phys. Rev. B \textbf{54}, 14679 (1996).

\bibitem{ryazanov_prl_01} V. V. Ryazanov \etal, Phys. Rev. Lett. \textbf{86}, 2427 (2001); T. Kontos \etal, Phys. Rev. Lett. \textbf{86}, 304 (2001).

\bibitem{khaire_prl_10} T. S. Khaire \etal, Phys. Rev. Lett. \textbf{104}, 137002 (2010).

\bibitem{sprungmann_arxiv_10} D. Sprungmann \etal, arXiv:1003.2082.

\bibitem{robinson_science_10} J. W. A. Robinson \etal, Science, 10 June 2010 (10.1126/science.1189246).

\bibitem{sperstad_prb_08} I. B. Sperstad \etal, Phys. Rev. B \textbf{78}, 104509 (2008)

\bibitem{houzet_prb_07} M. Houzet and A. I. Buzdin, Phys. Rev. B \textbf{76}, 060504 (2007).

\end{thebibliography}
\end{document}